\documentclass[twocolumn,pre,a4paper]{revtex4}
\usepackage{graphicx}
\usepackage{amsmath}
\newcommand{\ie}{{\em ie }}

\begin{document}
 LPT Orsay 02/105 
 
\title{\bf TREE NETWORKS WITH CAUSAL STRUCTURE}
\author{P. Bialas $^1$, Z. Burda $^2$, 
J. Jurkiewicz $^2$ and A. Krzywicki $^4$}
\affiliation{$^1$ Institute of Computer Science, 
Jagellonian University, ul. Nawojki 11, 30-072 Krakow, Poland\\
$^2$ M.~Smoluchowski Institute of Physics, Jagellonian University,
 ul. Reymonta 4, 30-059 Krakow, Poland\\
$^3$ Laboratoire de Physique Th\'eorique, B\^at. 210, 
Universit\'e Paris-Sud, 91405 Orsay, France\\}

\begin{abstract}
Geometry of networks endowed with a causal structure is
discussed using the conventional framework of equilibrium 
statistical mechanics. The popular growing network models
appear as particular causal models. We focus on a class 
of tree graphs, an analytically solvable case. General 
formulae are derived, describing the degree distribution, 
the ancestor-descendant correlation and the probability 
a randomly chosen node lives at a given geodesic distance 
from the root. It is shown that the Hausdorff dimension 
$d_H$ of the causal networks is generically infinite, in 
contrast to the maximally random trees, where it is 
generically finite.
\end{abstract}
\maketitle

\section{INTRODUCTION}
\subsection{Preamble}
   The network study is an old field of research, which has 
recently become particularly active (see \cite{rev}). This is 
largely due to the opening of an access to rich data on natural 
systems: the World-Wide Web, the Internet, the various biological 
networks (gene transcription, cell metabolism, protein 
interactions), the sociological networks (citation index, 
collaborations, phone calls), etc. Interesting empirical 
regularities have been observed, like the small-world 
property of networks or the frequently observed scale-free 
nature of degree distributions. On the theory side, the natural 
conceptual framework for network research is the graph theory: the 
nodes and the links of a graph represent the active agents and 
the interactions, respectively. However, it 
was soon realized that the classical 
graph theory \cite{bol} is inadequate and has to be generalized, in order 
to cover the new reality. This observation has triggered, in 
turn, an intense theoretical activity, that led to the construction 
of a number of insightful models. It is clear, however, that 
much remains to be done.
\par
As recalled in ref. \cite{bck}, in studying complex systems 
one can adopt one of the two complementary approaches: the 
diachronic and the synchronic one. In the former, one focuses on
the time evolution of the system, which helps discovering
the dynamics at work. In the latter, one considers an ensemble 
of dynamically similar systems at a fixed large time, which helps 
identifying the generic structural traits.
\par
Most of the recently constructed models adopt the diachronic
approach and deal with growing networks 
\cite{ba}-\cite{blw}.
This led, among others, 
to the remarkable discovery of the preferential attachment rule 
and of its crucial role in network 
evolution \cite{ba,baj}. The mathematics of
the diachronic approach rests largely on master equations and related
manipulations. The synchronic approach has also been advocated 
and various static statistical ensembles have been constructed
and studied \cite{bck},\cite{nsw}-\cite{ffh}.
Here, the mathematics is that of the equilibrium 
statistical mechanics and probability theory. 
One of the aims of this paper is to
help establishing a bridge between the two sets of results. 
It will be seen that the widely accepted distinction 
between growing and equilibrium networks is not really
correct. The truly distinctive property of growing networks,
their causal structure, can be usefully incorporated in a
static model. We limit
ourselves to tree graphs, because in this case the use of the
methods of analytic combinatorics enables one to get
exact solutions.
\par
We shall discuss at length the geometry of connected tree 
graphs endowed with a causal structure. Our discussion
will cover, in particular, some of the recently popular
growing network models \cite{ba,kr}, but we shall proceed 
considering static statistical ensembles of trees, 
employing a formalism which proved to be useful in deriving 
generic results for other random geometries. For example, we 
shall show that the causal trees have the "small world" property: their 
Hausdorff dimension is infinite. The advantage of our 
approach is that is enables one to show at once that this property 
holds in a wide class of models. 
\par
The plan of the paper is as follows: In the next two subsections
we recall some basic concepts and we define 
the observables to be calculated later on. In sect. II we 
begin by deriving a number of general results that 
hold for causal trees: The central role is played by a recursion 
relation satisfied by the partition function (sect. IIA). This 
recursion relation enables one to find closed expressions for the 
degree distribution (sect. IIB), the ancestor-descendant correlation 
(sect. IIC) and the two-point function measuring the distribution of 
shortest path lengths between pairs of nodes (sect. IID). In 
sect. IIE we establish contact between our formalism and the growing 
network models. This is used in sect. IIF, where we calculate 
explicitly a set of observables for some simple growing networks. 
In sect. III we compare the causal model with the maximally 
random one: In sect. IIIA we collect all general formulae. In 
sect. IIIB we calculate the observables, assuming the same 
microstate weights as in sect. IIF, and we find dramatically 
different results. This illustrates the importance of graph 
symmetries. In sect. IIIC we show how to construct a maximally 
random model with the same degree distribution as a given causal 
model. We stress, however, that the identity of degree distributions 
does not imply that geometries are similar. In particular, the average shortest 
path lengths are very different: generically they scale like a power 
of the number of nodes $N$ in maximally random and like 
$\log{N}$ in causal models, respectively. We conclude briefly 
in sect. IV.

\subsection{Basic concepts}
Let us recall some definitions: A {\em rooted} tree is a tree with
one marked node. A {\em planted} tree is a rooted one with an extra
link attached to the root, so that the degree of the root is
increased by unity. The other end of this extra link is
not counted as a node, in a sense this end remains "free". 
The different tree ensembles are simply related and choosing 
to work with one of them is a matter of convenience. In this 
paper we deal with planted tree graphs \cite{polya}.
\par
By assumption, a label is attached to each node and two graphs
with identical topology but labeled differently are considered
different. We say a tree is endowed with a {\em causal
structure} when the labels always appear in growing numerical order
as one moves along the tree from the root towards an arbitrary
node. These are the tree graphs we are interested in. 
\par
Introduce a statistical ensemble of these trees. Let us denote 
by $T$ a given topology and by $L(T)$ the number of distinct causal 
labelings of $T$. We attach the same weight
\begin{equation}
\rho(T)= w(n_1, n_2, \ldots , n_N)
\end{equation}
to each acceptable labeling. Here $N$ is the total 
number of nodes, $n_i$ denotes the degree of the node $i$ and 
$w$ is some appropriate positive-definite function. A 
model is defined by choosing a particular form of $w$.
 
\par
It will be seen that the presence of a causal structure 
generates non-trivial observable inter-node correlations. 
Hence, it is of interest to discuss models where these 
specific correlations do not interfere with correlations of 
a different origin. With this motivation, we assume in this paper
that $\rho(T)$ factorizes \cite{foot0}
\begin{equation}\label{eq:prod}
\rho(T)=\prod_{i=1}^N q_{n_i}
\end{equation}
On the other hand, we keep $q_n$ as general as possible.
\par
The partition functions of canonical and grand-canonical 
ensembles are defined in the usual way by summing the weights of
all possible microstates. Thus by definition~:
\begin{equation}\label{eq:zn}
z_N=\frac{1}{N!}\sum_T L(T) \rho(T)
\end{equation}
and
\begin{equation}\label{eq:zlambda}	
Z(\mu)=\sum_N z_N e^{-\mu N}
\end{equation}
\par
The prefactor $1/N!$ in eq. \eqref{eq:zn} is compensated
by the number of terms in the summand. Indeed, the
number of labeled causal trees with $N$ nodes is $(N-1)!$ 
\cite{foot1}. It is not difficult to convince oneself that
in all cases of physical interest $z_N$ grows exponentially
with $N$, up to a power prefactor~:
\begin{equation}\label{eq:lambdac}
z_N \sim e^{\bar{\mu} N} \; \;  \mbox{\rm for} \; \; N \to \infty
\end{equation}
and $Z(\mu)$ develops a singularity at $\mu=\bar{\mu}$ (see sect. IIA
for a more rigorous argument). One is 
primarily interested in the regime controlled by
the singular part of $Z(\mu)$. Indeed, as 
$\delta\mu \equiv \mu - \bar{\mu}$ tends to zero one becomes 
increasingly sensitive to the behavior of trees with arbitrarily 
large $N$.
\par
Of course, eq. \eqref{eq:zlambda} can be inverted by Laplace
transformation
\begin{equation}\label{eq:znlaplace}	
z_N = \frac{e^{\bar{\mu} N}}{2\pi i}\int_{-i\infty}^{i\infty} 
\text{d}\delta\mu \; Z(\delta\mu)\; e^{\delta\mu N}
\end{equation}
with the integration contour passing on the right of the
singularity at $\delta\mu=0$. It will often be convenient 
to work in the grand-canonical ensemble and to Laplace 
transform the result to the physically more interesting 
canonical ensemble at the very end. In practice, we shall always
assume that $N \to \infty$ and we shall keep the leading term only.
\par
Notice, that there is an analogy between the labelings of 
graphs and the positions of a system in a (discrete) phase-space. 
Hence, $\rho(T)$ and $L(T)$ are the analogues of the weight of a 
microstate and of the corresponding phase-space volume, respectively.
\par
Little is specific to causal trees in the content of this section.
One could repeat {\em verbatim} the above definitions and keep the 
same microstate weights in the context of a different tree ensemble. 
However, in this new ensemble the trees would have different symmetries 
and, therefore, $L(T)$ would be in general different. Consequently, 
the physics would be different too. We shall see later, comparing 
causal and maximally random trees, that introducing a new symmetry 
can change dramatically the geometry of generic graphs.

\subsection{Observables}
Let us define the observables we shall calculate in this paper~:
\par
The most popular observable is the {\em degree distribution}. When the 
weight of a microstate has the factorized form, as in \eqref{eq:prod}, 
the degree distribution $\pi_n$ is given by the simple and obvious 
formula
\begin{equation}\label{eq:deg}
\pi_n = N^{-1}q_n \frac{\partial \ln{z_N}}{\partial q_n}
\end{equation}
The factor $N^{-1}$ above is included to have the distribution
normalized to unity. 
\par
In the thermodynamic limit $N \to \infty$
\begin{equation}\label{eq:degree}
\pi_n = q_n \frac{\partial \bar{\mu}}{\partial q_n}
\end{equation}
\par
The next observable is the {\em correlation} between node degrees, say the 
probability that a node has degree $k$ when its neighbor's degree is $l$.
In a causal tree, one of these nodes is an {\em ancestor} and the other a
{\em descendant}. 
\par
A very interesting observable is the {\em Hausdorff dimension} 
$d_H$ controlling the scaling with $N$ of the linear size of a 
typical tree~:
\begin{equation}\label{eq:fractal}
\langle r \rangle_N \sim N^{1/d_H}
\end{equation}
One usually takes for $r$ the distance between an arbitrary pair of nodes. 
In rooted trees it is more natural to consider the distance separating 
a randomly chosen node from the root. One first calculates a specific 
two-point function $C(r, \mu)$, the grand-canonical weight of all 
trees with a node separated from the root by $r$ steps. Using 
$C(r, \mu)$ one finds 
\begin{equation}
\langle r \rangle_\mu = \frac{\int_0^{\infty} \text{d}r \; r\;  C(r, \mu)}
{\int_0^{\infty} \text{d}r \;  C(r, \mu)}
\end{equation}
This quantity usually diverges when $\delta\mu \to 0$. The 
behavior on the r.h.s. (right-hand side) of \eqref{eq:fractal} is 
determined by observing that $\delta\mu$ scales like $N^{-1}$ 
(see \eqref{eq:znlaplace}).

\section{GEOMETRY OF TREES ENDOWED WITH A CAUSAL STRUCTURE}
\subsection{Recursion relation}
We start by deriving a recursion relation for the partition function
$z_N$. To this end we construct a new planted tree by attaching the $k$
"free link-ends" of the planted trees $T_1,\ldots,T_k$ to a new root.
We denote the resulting compound tree by $T=T_1\oplus \cdots\oplus T_k$ 
(this is illustrated in FIG. 1 for $k=3$).
\begin{figure}
\includegraphics[width=8cm]{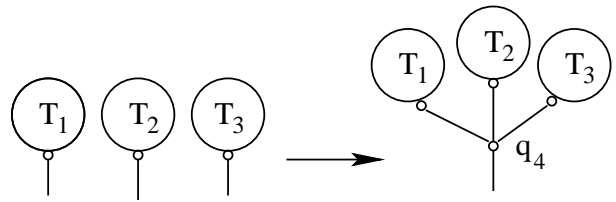}
\caption{ Construction of the compound tree $T_1\oplus T_2\oplus T_3$.
The three old roots are connected to the new root with degree four. Notice
that all the graphs are planted trees. The trees are not planar and
therefore the relative position of branches in the compound tree is 
irrelevant. }
\end{figure}
\par
The number of ordered labelings of the compound tree is 
\begin{equation}\label{eq:L}	
L(T_1\oplus \cdots \oplus T_k)=
\frac{N!}{N_1!\cdots N_k!}\frac{1}{k!}L(T_1)\cdots L(T_k)
\end{equation}
Here $N_i$ denotes the total number of nodes in the tree $T_i$ 
and $N=\sum_i N_i$. One has to give $N+1$ labels to the nodes of the 
compound tree. However, the smallest label must be attached to the root. 
The remaining $N$ labels are arbitrarily distributed among the trees. 
This is the origin of the the multinomial factor. Permuting the trees 
$T_i$ does not change the compound tree \cite{foot2}. This explains 
the presence of the factor $1/k!$. It will be seen that the {\em 
composition rule} \eqref{eq:L} is a very basic property of the model.
\par
Obviously the weight of the new tree factorizes
\begin{equation}\label{eq:rho}	
\rho(T_1\oplus \cdots \oplus T_k)=q_{k+1}\rho(T_1)\cdots \rho(T_k)
\end{equation}
\par
The partition function $z_{N+1}$ can be constructed by summing the
trees of size smaller or equal to $N$~:
\begin{equation}\label{eq:ZN}\begin{split}	
z_{N+1}=&\frac{1}{(N+1)!}\sum_{k=1}^\infty 
\sum_{T_1,\ldots,T_k}\delta_{N_1+\cdots+N_k,N}\\
& \times L(T_1\oplus\cdots\oplus T_k)\rho(T_1\oplus\cdots\oplus T_k)
		\end{split}
\end{equation} 
\par
Inserting \eqref{eq:L} into \eqref{eq:ZN} and rearranging terms in
the sum we obtain after trivial algebra~:
\begin{equation}\label{eq:rec}
z_{N+1} = \frac{1}{N+1}\sum_{k=1}\frac{q_{k+1}}{k!}
\!\sum_{N_1,\ldots,N_k}\kern-3mm\delta_{N_1+\cdots+N_k,N}\prod_{i=1}^kz_{N_i}
\end{equation}
Adding $z_1=q_1$ and summing both sides of equation \eqref{eq:rec} we get
\begin{equation}	
\sum_N N z_N e^{-N \mu}=
e^{-\mu}\left(\sum_{k=0}^{\infty} \frac{q_{k+1}}{k!}Z(\mu)^k\right)		
\end{equation}
or finally 
\begin{equation}\label{eq:diff}
Z'(\mu) = - \, e^{-\mu} F(Z)
\end{equation}
where 
\begin{equation}\label{eq:defF}
F(Z) = \sum_{k=1}^{\infty} \frac{q_k}{(k-1)!} Z^{k-1}
\end{equation}
Equation \eqref{eq:diff} can be integrated to give
\begin{equation}\label{eq:Z}	
e^{-\mu(Z)} = G(Z) \equiv \int_0^Z\frac{\text{d}x}{F(x)}
\end{equation}
The function $G(Z)$ is a positive monotonically growing function of
Z, bounded from above (one can ignore the trivial case where all 
$q_n$ except $q_1$ and $q_2$ are zero). Hence $\mu$ is bounded
from below: $Z(\mu)$ has a singularity at some $\mu=\bar{\mu}$.
Denote by $\bar{x}$ the radius of convergence of the series $F(Z)$.
The critical value of $\mu$ is given by
\begin{equation}\label{eq:muc}
\bar{\mu} = -\log G(\bar{x})
\end{equation}
This formula holds also when the radius of convergence $\bar{x}$ is infinite,
since all terms in the series \eqref{eq:defF} are positive and the
integral in \eqref{eq:muc} is convergent in all cases of interest~: 
$G(\infty)<\infty$. 

\subsection{Degree distribution}
The degree distribution calculated using \eqref{eq:degree} is  
\begin{equation}\label{eq:vdd}
\pi_n = \frac{1}{G(\bar{x})}\frac{q_n}{(n-1)!}
\int_0^{\bar{x}}\kern-2mm\frac{\text{d}x}{F(x)^2}x^{n-1}		
\end{equation}
Again, this formula is also valid when $\bar{x}=\infty$. 
\par
Summing over $n$ and using the definitions of $F$ and $G$ one easily
checks that $\pi_n$ is normalized to unity, as it should. One further 
finds
\begin{equation}\label{eq:sumto2}
\sum_n n \; \pi_n = 2 - \frac{\bar{x}}{F(\bar{x}) G(\bar{x})}		
\end{equation}
On a tree, the r.h.s. should equal 2 . This is the case when $F(x)$ 
diverges at $x=\bar{x}$. Otherwise one encounters a pathology, which looks
similar to that appearing in some maximally random tree models (and in 
the so-called balls-in-boxes model, see \cite{bbj,bck}), where working
in the large $N$ limit one misses singular node(s) contributing term(s) 
of the type $N^{-1} \delta (n-cN)$. In the limit such nonuniformly 
behaving terms do not contribute to the normalization but do contribute 
to the r.h.s. of \eqref{eq:sumto2}. It will be shown later that when
$F(\bar{x}) < \infty$ the average distance between nodes is finite. This  
means that singular node(s) -  with unbounded connectivity - 
are indeed expected to show up. In deriving \eqref{eq:vdd} the large 
$N$ limit has been implicitly used and it is an educated guess 
that one again misses the singular node(s).

\subsection{Ancestor-descendant correlation}
Now, we turn to the calculation of the ancestor-descendant degree 
correlation. It is obvious that an ancestor plays the role of the root
of the subgraph involving all its descendants. One can read from
\eqref{eq:rec} the degree distribution of the root:
\begin{equation}
z_l(N) = \frac{1}{N}\frac{q_l}{(l-1)!} 
\kern-1mm\sum_{N_1,\dots,N_{l-1}}\kern-3mm \delta_{N_1+\cdots+N_{l-1}, N-1} 
\prod_1^{l-1} Z_{N_i}
\end{equation}
Going over to the grand-canonical ensemble one finds:
\begin{equation}
\frac{\text{d}Z_l(\mu)}{\text{d}\mu} = 
-e^{-\mu} \frac{q_l}{(l-1)!} Z^{l-1}(\mu)
\end{equation}
which, taking \eqref{eq:diff} into account and after integration yields
\begin{equation}
Z_l\bigl(\mu(Z)\bigr) =  \frac{q_l}{(l-1)!} 
\int_0^Z \text{d}x \frac{x^{l-1}}{F(x)}
\end{equation}
Using similar arguments one writes the weight of graphs where the 
root has the degree $l$ and its daughter the degree $k$ as
\begin{equation}\begin{split}
z_{kl}(N) = &\frac{q_l}{N (l-2)!}\sum_{N_1,\dots,N_{l-1}} 
\kern-3mm\delta_{N_1+\dots+N_{l-1}, N-1} \times\\
&\times \prod_{i=1}^{l-2} Z_{N_i} z_k(N_{l-1})
\end{split}
\end{equation}
Hence
\begin{equation}
\frac{\text{d}Z_{kl}(\mu)}{\text{d}\mu} = -e^{-\mu} 
\frac{q_l}{(l-2)!} Z^{l-2}(\mu)Z_k(\mu)
\end{equation}
Integrating the above equation one finally obtains 
\begin{multline}\label{eq:2ptcor}
Z_{kl}\bigl(\mu(Z)\bigr) = \frac{q_l}{(l-2)!} \frac{q_k}{(k-1)!}\times \\
\times \int_0^Z \text{d}x_2 \frac{x_2^{l-2}}{F(x_2)} 
\int_0^{x_2} \text{d}x_1 \frac{x_1^{k-1}}{F(x_1)}
\end{multline}
which is the conditional probability, up to normalization, that a 
descendant has the degree $k$ when the ancestor's degree is $l$.
The normalization is determined summing over $k$ on the 
r.h.s. above, with the result $(l-1)Z_l(\mu)$. It is a
slightly different measure of neighbor correlation than that
proposed in ref. \cite{kr}, but it carries similar information.
Because of the integration the dependence on $k$ and $l$ does 
not factorize, in general. 
\par
Eq. \eqref{eq:2ptcor} holds in the grand-canonical 
ensemble. One would like to have an expression valid in the 
canonical ensemble, where the graph has a well defined number 
of nodes. This requires Laplace transforming the r.h.s. of 
eq. \eqref{eq:2ptcor} (one should transform first 
and normalize next). One can argue that in the thermodynamical
limit and in the regime $\delta\mu \to 0$ this often amounts 
to just replace $Z \to \bar{x}$ in \eqref{eq:2ptcor}. Indeed, let   
$Z-\bar{x} \sim \delta\mu^\alpha$. Then
\begin{equation}
\int \text{d}\delta\mu \; e^{\delta\mu N}\; Z_{kl}(\mu) \approx
Z_{kl}(\bar{\mu}) \int \text{d}\delta\mu\; e^{\delta\mu N - c\, 
\delta\mu\,^{\alpha}}
\end{equation}
Evaluating the integral by the saddle-point method one gets an
exponential of a quantity scaling like $N^{\alpha/(\alpha-1)}$.
This gives a factor of unity in the limit $N \to \infty$, provided
$0 < \alpha < 1$, a condition often met in applications. 

\subsection{Fractal dimension}
Repeating over and over  the iteration process 
leading to eq. \eqref{eq:2ptcor} one gets
\begin{multline}\label{eq:rptcor}
Z_{k_1k_2 \dots k_r}(\mu(Z)) =   \\
 \prod_{j=2}^r \frac{q_{k_j}}{(k_j-2)!}\; \frac{q_{k_1}}{(k_1-1)!}
\int_0^Z \text{d}x_r \frac{x_r^{k_r-2}}{F(x_r)} \times  \\
\times \int_0^{x_{r}}\kern-3mm\text{d}x_{r-1} 
\frac{x_{r-1}^{k_{r-1}-2}}{F(x_{r-1})} 
\cdots \int_0^{x_2} \text{d}x_1 \frac{x_1^{k_1-1}}{F(x_1)}\; \; \;
\end{multline}
Summing over node degrees $k_1, k_2, \cdots , k_r$ one obtains 
the weight of all graphs with a point separated by $r$ links from 
the root, \ie the two-point correlation
function $C(r, \mu)$ introduced in sect. 1.2~:
\begin{multline}
C\bigl(r, \mu(Z)\bigr) = \\\int_0^Z \text{d}x_r \frac{F'(x_r)}{F(x_r)} 
\int_0^{x_r} \text{d}x_{r-1} \frac{F'(x_{r-1})}{F(x_{r-1})}\cdots \times\\
\times \int_0^{x_3} \text{d}x_2 \frac{F'(x_2)}{F(x_2)}
\int_0^{x_2} \text{d}x_1 \; \; \; \;
\label{cr} 
\end{multline}
For finite $\bar{x}$, replacing the upper limit 
of integration over $x_1$ by $\bar{x}$ and
performing all integrations, one gets
\begin{equation}
C\bigl(r, \mu(Z)\bigr) \leq \bar{x} \frac{\bigl(\ln F(Z)\bigr)^{r-1}}{(r-1)!}
\end{equation}
Hence, the tail of $C(r, \mu)$ falls at least as fast as a Poissonian.
Consequently $\langle r \rangle_\mu$ grows at most like $\ln{F(Z)}$.
Assuming that $F(z)$ has at most a power singularity 
at $z=\bar{x}$ one concludes that  
\begin{equation}
\langle r \rangle_\mu  \leq 
\mbox{\rm const} \ln{\frac{1}{\delta\mu}}
\end{equation}
and therefore
\begin{equation}
\langle r \rangle_N \leq \mbox{\rm const} \ln{N}
\end{equation}
since $\delta\mu$ scales like $N^{-1}$. The argument is rather heuristic, 
but suggestive (see also the examples in the next section). 
It appears that generically the causal trees have the {\em small-world} 
property $d_H=\infty$, contrary to the maximum entropy trees 
whose generic fractal dimension is finite \cite{adj,jk,bck}.
This phenomenon is easy to understand intuitively~: the causal structure 
suppresses long branches. This can be seen by noting that along 
a branch from the root to the leaf no label permutations are possible, 
hence a tree with a few long branches admits much less causal labelings  
then a ``short fat'' one.

\subsection{Synchronic view of growing networks}
So far, our discussion was very general. Let us now 
establish a bridge to the popular growing network
models. Of course, here we consider only those models,
where one constructs tree graphs. Successive nodes are
attached, one at a time, the attachment probability being 
a function of the degree of the target node~:
\begin{equation}	
\text{Prob}(k\mid t) = \frac{A_k}{A(t)}
\quad\text{with}\quad A(t)=\sum_k N_k(t) A_k
\end{equation}
where $N_k(t)$ is the number of vertices with degree $k$ in the tree
at time $t$. It is obvious that the tree constructed that way has a
causal structure: nodes are labeled by the attachment time.
\par
Notice, that from the perspective of a model builder, the
concepts of causal and growing networks are 
complementary rather than equivalent.
A causal network is defined by specifying the microstate weights.
Every growing network is causal, of course, but the weights
corresponding to a given growth process can have a very complicated,
non-local structure. And conversely, given a set of
weights it is, in general, not evident what is the corresponding 
growth process. Only for a class of models there exists a stationary
attachment kernel $A_k$.
\par
For linear or shifted linear attachment kernels $A_k$ the normalization
factor $1/A(t)$ depends only on the size of the previous configuration and 
is therefore the same for all trees of the same size \cite{foot3}. Hence,
the preferential attachment recipe is compatible with the factorization 
of $\rho(T)$: Working in the canonical ensemble, at fixed time, we can 
drop the normalization factor altogether without any loss of generality 
and set \cite{foot4}
\begin{equation}	
q_n=q_1\prod_{k=1}^{n-1}A_k,\quad n>1\quad
\end{equation}
where $q_1$ is some positive constant (eventually set to 1 in explicit
calculations). It is instructive to check by inspection that 
the graph weights produced 
by the recursion relation \eqref{eq:rec}, coincide with 
those generated by the growing network model recipe.

\subsection{Examples}
\subsubsection{Barabasi-Albert model \cite{ba}}
In this model $A_k=k$. Therefore $q_n=(n-1)!$,
$F(x)=(1-x)^{-1}$ and $\bar{x}=1$. Thus $G(x)=x-\frac{1}{2}x^2$
and the solution to the equation \eqref{eq:Z} is 
\begin{equation}	
Z(\mu)=1-\sqrt{1-2e^{-\mu}} \approx 1-\sqrt{\delta\mu}
\end{equation}
The degree distribution is found from \eqref{eq:vdd}~: 
\begin{equation}
\pi_n=2 \int_0^{1}\text{d}x\;(1-x)^2\;x^{n-1}
=\frac{4}{n(n+1)(n+2)}
\end{equation}
which coincides with the solution given in refs. \cite{kr,dm}.
\par
The near-neighbor correlation is readily found from 
\eqref{eq:2ptcor}~:
\begin{eqnarray}
Z_{kl} &=& (l-1) \int_0^1\kern-2mm \text{d}x_2\; (1-x_2)\; x_2^{l-2} 
\int_0^{x_2}\kern-3mm\text{d}x_1 \; (1-x_1)\;x_1^{k-1} \nonumber \\
&= & \frac{l-1}{k(k+1)(k+l-1)(k+l)} \; \; \; 
\end{eqnarray}
This is very similar to the result of ref. \cite{kr}, where a 
slightly different quantity has been calculated. The physical 
content is the same: the causal structure has induced correlations
between the node degrees.
\par
Since in this example
\begin{equation}
\int_0^{x_3} \text{d}x_2 \frac{F'(x_2)}{F(x_2)} \int_0^{x_2} \text{d}x_1 =
\int_0^{x_3} \text{d}x_2 [\frac{F'(x_2)}{F(x_2)}- 1]
\end{equation} 
one has
\begin{equation}
C\bigl(r, \mu(Z)\bigr) = 
\frac{\bigl(\ln F(Z)\bigr)^{r-1}}{(r-1)!}- C\bigl(r-1, \mu(Z)\bigr)
\end{equation}
Hence
\begin{equation}
\langle r \rangle_\mu \sim \ln F(Z) \sim \ln{\frac{1}{\sqrt{\delta\mu}}}
\end{equation}
which implies the following scaling law~:
\begin{equation}
\langle r \rangle_N \sim \frac{1}{2}\ln{N}
\end{equation}

\subsubsection{Krapivsky-Redner model \cite{kr}}
Now $A_k=k+w$, $q_n=\Gamma(n+w)$ and  
\begin{equation}	
F(x) = \Gamma(w+1)(1-x)^{-w-1}.
\end{equation}
Of course $\bar{x}=1$ and
\begin{equation}
G(1)= \frac{1}{(2+w)\Gamma(w+1)}
\end{equation}
Using the above and evaluating the Euler integral that
appears in the present case on the r.h.s. of \eqref{eq:vdd}
one obtains 
\begin{equation}\label{eq:krd}
\pi_n=\frac{(2+w)\Gamma(3+2 w)}{\Gamma(1+w)}
\frac{\Gamma(n+w)}{\Gamma(3+n+2w)}
\end{equation}
Again this reproduces exactly the result of ref. \cite{kr}.
The reader can easily calculate the correlation $Z_{kl}$.
We skip this calculation here, because the result is not 
particularly instructive: the remarkable fact is the very 
existence of the correlation, not its particular form, which 
in this particular case is rather cumbersome.
\par
The calculation of $C(r, \mu)$ is identical to that carried out
for the Barabasi-Albert model, except that
\begin{equation}
Z=1-\delta\mu^{\frac{1}{w+2}} \; , \; \; \delta\mu \to 0
\end{equation}
which implies
\begin{equation}
\langle r \rangle_N \sim \frac{w+1}{w+2}\ln{N}
\end{equation}

\subsubsection{Constant attachment kernel}
For $A_k=1$ 
\begin{equation}	
F(x)=\sum_{k=1}^\infty\frac{1}{\Gamma(k)}x^{k-1}=e^x,
\end{equation}
The interest of this example is in the infinite
radius of convergence of the above series: $\bar{x}=\infty$.
The degree distribution is 
\begin{equation}
\pi_n=
\frac{1}{\Gamma(n)}
\int\limits_0^\infty\text{d}x e^{-2x} x^{n-1}
=2^{-n}
\end{equation}
a result found by several people, including the
authors of ref. \cite{kr}. We again skip the calculation of
$Z_{kl}$. 
\par
Since $d\ln F(x) = dx$ one finds
\begin{equation}
C\bigl(r, \mu(Z)\bigr) = \frac{Z^r}{r!}
\end{equation}
and using
\begin{equation}
Z = \ln \frac{1}{\delta\mu} \; , \; \; \delta\mu \to 0
\end{equation}
one derives
\begin{equation}
\langle r \rangle_N \sim \ln{N}
\end{equation}

\section{CAUSAL VERSUS MAXIMALLY RANDOM TREES}
\subsection{An important tiny difference}
It is instructive to consider for also the well known case 
of maximum entropy trees (cf. refs. \cite{adj,bb,jk,bck}). 
This will help putting the results of the preceding section 
in proper perspective. As in the preceding section we start 
with the composition rule for trees, which in the present 
case, when the causality constraint is lifted, reads
\begin{equation}\label{eq:L'}	
L(T_1\oplus \cdots \oplus T_k)=\frac{(N+1)!}{N_1!\dots N_k!}
\frac{1}{k!}L(T_1)\cdots L(T_k)
\end{equation}
Compared to \eqref{eq:L} the difference may seem tiny: one just has
$(N+1)!$ instead of $N!$ in the numerator on the r.h.s.. It is so
because there is no causality constraint and therefore all $N+1$ 
labels can be arbitrarily distributed among the trees. However,
this tiny difference has rather dramatic consequences. Indeed,
repeating the steps which led from (\ref{eq:L}) to (\ref{eq:diff})
one obtains
\begin{equation}\label{eq:diff'}
Z(\mu)=e^{-\mu}F(Z)
\end{equation}
No derivative appears on the left-hand side! Instead of eq. 
\eqref{eq:Z} one has
\begin{equation}\label{eq:hz}
e^{-\mu} = H(Z) \equiv \frac{Z}{F(Z)}
\end{equation}
where the function $H(Z)$ plays the role analogous to that of
$G(Z)$ in the preceding section. Eq. \eqref{eq:hz} implies 
that the vertex degree distribution is
\begin{equation}\label{eq:newpin}
\pi_n=
\frac{1}{F(x_*)} \frac{q_n\, x_*^{n-1}}{(n-1)!}
\end{equation}
Here $x_* = \min( x_{max},\bar{x})$ and $x_{max}$ is the position
of the maximum of the function $H(x) : H'(x_{max})=0,\; H''(x_{max})<0$. 
\par
Since no derivatives appear as one goes over to the grand-canonical
ensemble, one does not integrate either. Therefore, iterating the 
recursion relation one finds that $Z_{kl}(\mu)$ and $C(r, \mu)$ 
factorize. The latter equals
\begin{equation}
C(r, \mu) \propto \bigl(e^{-\mu} F'(Z)\bigr)^r
\end{equation}
as first derived by Ambj\o rn et al \cite{adj} using diagrammatic
arguments.
\par
Let us now consider three examples, with $q_n$ chosen as in sects.
IIF.1 - IIF.3 respectively. 

\subsection{Examples}
\subsubsection{First example: $q_n = (n-1)!$}
With this choice, that of sect. IIF.1, the function $H(x)$ reads 
$H(x) = x - x^2$ and has its maximum at $x=1/2$. Hence $x_*=1/2$. 
Since $F(x_*)=2$, the degree distribution calculated from 
\eqref{eq:newpin} is now
\begin{equation}
\pi_n= 2^{-n}
\end{equation}
as in sect. II.F.3 !
\par
It is easy to check that for small $\delta\mu$ one has
\begin{equation}
e^{-\mu} F'(Z(\mu)) = 1 - \sqrt{\delta\mu} 
\end{equation}
and therefore
\begin{equation}
C(r, \mu) \propto e^{-\sqrt{\delta\mu}\, r} 
\end{equation}
which implies
\begin{equation}
\langle r \rangle_N \sim N^{1/2} 
\end{equation}
Hence, $d_H=2$, the generic value \cite{adj}.

\subsubsection{Second example: $q_n = \Gamma(n+w)$}
With the choice of sect. IIF.2 the function 
$H(x) = \Gamma(w+1) x (1-x)^{w+1}$ has its maximum
at $x_{max} = 1/(2+w)$, which is smaller than the radius
of convergence $\bar{x}=1$ of the series $F(x)$ as long as $w>-1$.
Thus $x_*=1/(2+w)$ and 
\begin{equation}
\pi_n=  \frac{2+w}{\Gamma(1+w)} \left(\frac{1+w}{2+w}\right)^{1+w} 
 \frac{\Gamma(n+w)}{\Gamma(n)} \, \left(\frac{1}{2+w}\right)^n
\end{equation}
The fall of the degree distribution is again exponential. One
easily checks that again $d_H=2$.

\subsubsection{Third example: $q_n = 1$}
With the choice of sect. IIF.3 the function $H(x) =  x e^{-x}$
has its maximum at $x_{max}=1$. Thus $x_*=1$ and correspondingly
\begin{equation}
\pi_n = \frac{e^{-1}}{(n-1)!}
\end{equation}
One can easily check that this result holds for
a more general family of weights $q_n = \alpha \beta^{n-1}$ 
independently of the values of $\alpha$ and $\beta$, 
as long as they are positive. Again $d_H=2$.

\subsection{How to get identical degree distribution in the two models}
It is easy to adjust the input parameters of the two models 
to obtain identical degree distributions: Suppose that in 
the causal model the degree distribution is $\pi_n$. For 
a connected tree one necessarily has
\begin{equation}\label{eq:pin}
\sum_n n \pi_n  = 2
\end{equation}
Set in the maximally random model
\begin{equation}
q_n = (n-1)! \; \pi_n
\end{equation} 
Obviously 
\begin{equation}
[H^{-1}]'(1) \equiv \sum_n (n-2)\pi_n = 0
\end{equation} 
and $[H^{-1}]''(1) > 0$. Thus, $H(x)$ has a maximum at $x=1$.
One easily convinces oneself that this is the only maximum of
that function. Furthermore, by assumption (see \eqref{eq:pin}),
the radius of convergence in $F(x)$ is  $\geq 1$. Hence $x_*=1$ and
\begin{equation}
\pi_n^{\mbox{\footnotesize \rm random}} = \pi_n
\end{equation}
by virtue of eq. \eqref{eq:newpin}. However, the coincidence 
of the degree distributions does not imply that geometries are 
similar. On the contrary, as already emphasized, the graph linear sizes
have generically a very different scaling behavior in 
the two models (see also some computer simulation 
results in \cite{krz}).

\section{SUMMARY AND CONCLUSION}
The results of this paper are another illustration of the claim 
that the opposition between diachrony and synchrony is to large 
extent an illusion, except if one is interested in very specific
phenomena, like aging, intrinsically reflecting the running of
time. We have discussed the geometry of networks endowed with a causal
structure using the conventional framework of equilibrium statistical
mechanics. Hence, models that are usually described by specific
master equations and static, maximum entropy models have been treated
alike. We focused on tree graphs, because only for trees we are 
able to proceed analytically. We have derived general formulae 
describing the degree distribution, the ancestor-descendant 
correlation and the probability a node lives at a given geodesic 
distance from the root. Using this last results have shown that 
our causal networks have generically the small-world property, 
\ie their Hausdorff dimension is infinite. 
\par
We have also compared the causal model with the maximally random one, 
assuming the same microstate weights. Because of different symmetry 
properties of the graphs - in the causal model only a subclass of 
labelings is allowed - the geometries are dramatically different: 
in the causal model the degree distribution is qualitatively different, 
inter-node correlations are induced and the Hausdorff dimension 
becomes infinite instead of being finite.

{\bf Acknowledgments} This work was partially supported by 
the EC IHP network HPRN-CT-1999-000161 and by grant 2 P03B
096 22 of the Polish Research Foundation (KBN) in years 2002-2004. 
Laboratoire de Physique Th\'eorique is Unit\'e Mixte du CNRS UMR 8627.


\begin{thebibliography}{99}
\bibitem{foot1} The structure of a causal tree is determined if one
gives the labels of the parents of each node. The root has label 1 and 
no parent. The node with label 2 has only one parent, the root. There are
two candidates for being the parent of the node with label 3: the root and
the node with label 2. There are three candidates for being the parent of 
the node with label 3, etc. Pursuing the argument and counting the number 
of distinct trees one obtains the result quoted in the text.
\bibitem{foot3}For other kernels one could perhaps make the approximation 
done by the authors of ref. \cite{kr}, \ie assume that $A(t)=\mu N$.
\bibitem{foot4} Those factors may be important if one wants 
to consider the thermodynamic limit $N\rightarrow\infty$.
\end{thebibliography}
\end{document}